\documentclass[preprint,pre,showkeys,showpacs]{revtex4}
\usepackage{epsfig}
\usepackage{amsmath}
\begin{document}
%\title{Clustering, Crisis and Dynamical Regimes in a Bailout Embedding Map}
%title{Clustering, Crisis and the Dynamics of Inertial Particles of a Bailout Embedding Map}
%\title{Clustering, Mixing and Crisis in a Bailout Embedding Map}
%\title{ An Attractor Merging Crisis in a Bailout Embedding Map}
\title{Clustering, Chaos and Crisis in a Bailout Embedding Map}
\author{N.Nirmal Thyagu}
\email{nirmal@physics.iitm.ac.in}
\author{Neelima Gupte}
\email{gupte@physics.iitm.ac.in}
\affiliation{Indian Institute of Technology Madras, Chennai-600036,India}
\date{\today}
\begin{abstract}
\indent We study the dynamics of inertial particles in two dimensional
incompressible flows. The particle dynamics is modelled by four
dimensional dissipative bailout embedding maps of the base flow which is
represented by $2-d$ area preserving maps. The phase diagram of the
embedded map is rich and interesting both in the aerosol regime, where
the density of the particle is larger than that of the base flow, as
well as the bubble regime, where the particle density is less than that
of the base flow. The embedding map shows three types of dynamic
behaviour, periodic orbits, chaotic structures and mixed regions.
Thus, the embedding map can target periodic orbits as well as chaotic
structures in both the aerosol and bubble regimes 
at certain values of the dissipation parameter.
The bifurcation diagram of the $4-d$ map is useful for the
identification of regimes where such structures can be found. An
attractor merging and widening crisis is seen for a special region  for
the aerosols. At the crisis, two period-10 attractors merge and widen
simultaneously into a single chaotic attractor. Crisis 
induced intermittency is seen at some points in the phase diagram.
The characteristic times before bursts at the
crisis show power law behaviour as functions of the dissipation
parameter. Although the bifurcation diagram for the bubbles looks similar to that of aerosols, no such crisis regime is seen for the bubbles. Our results can have  implications for the dynamics of impurities in diverse application contexts.
\end{abstract}

\maketitle

\section{Introduction}
	
Inertial particle dynamics, i.e. the dynamics of small spherical
particles immersed in fluid flows, has attracted considerable 
attention in  recent years\cite{mot03,reig01,babia00,cart02,benc03,raf06}.
These studies have two fold importance. From the point of view
of fundamental physics, the motion is governed by dynamical 
equations which exhibit rich and complex behaviour. From the 
point of view of applications, this dynamics constitutes
the simplest model for impurities whose transport in flows is of practical 
interest, in contexts as varied as the behaviour of aerosols or
pollutants in air, with consequences for weather and climate, and
that of nutrients such as plankton in the sea with consequences for 
ocean life.  

The Lagrangian dynamics of  such small spherical tracers in  two dimensional
incompressible fluid flows is described by the Maxey-Riley equations \cite{maxey83}.
These are further simplified under various approximations to give a 
set of minimal equations called the embedding equations where the fluid
flow dynamics is embedded in a larger set of equations which include the
the differences between the particle and fluid
velocities\cite{mot03,babia00}. 
Although the Lagrangian dynamics of the underlying fluid flow is
incompressible, the particle motion is  compressible \cite{maxey87},            and has regions of contraction and expansion. The density
grows in the former giving rise to clusters and
falls in the latter giving rise to voids. The properties of the base        
flow have important consequences for the transport and mixing of
particles. Map analogs of the embedding equations have also been 
constructed for cases where
the fluid dynamics is modelled by area-preserving maps which
essentially retain the qualitative features of the flow
\cite{pierrehumbert00,fereday02}.
The embedded  dynamics in both cases is dissipative in nature.

Further complexity is added to the dynamics by the density difference
between the particles and the fluid. In the case of two-dimensional
chaotic flows, it has been observed earlier that particles
with density higher than the base flow, the aerosols,
tend to migrate away from the KAM islands, 
while particles                                               
lighter than the fluid, the bubbles, display the opposite tendency
\cite{cartprl02}. Neutrally buoyant particles also showed a similar
result,  
wherein the particles settled into         
the KAM islands. Our study indicates that in addition to the density 
difference, the dissipation parameter of the system also has
an important role to play in the dynamic behaviour of the aerosols and       
the bubbles.           

In this paper, we study the dynamics of passive, finite size,
inertial particles in underlying
two dimensional incompressible flows. The dynamics is modelled by the
four dimensional dissipative embeddings of two dimensional area-preserving 
maps. We obtain  the phase diagram of the system in the $\alpha-\gamma$
space where $\alpha$ is the mass ratio parameter, and $\gamma$ is the
dissipation parameter. 
The phase diagram shows rich structure in the 
$\alpha < 1$ aerosol regime, as well as the $\alpha > 1$ bubble
regime.
Unlike the earlier results mentioned above, that the bubbles tended to form structures 
in the KAM islands\cite{cartprl02} and the aerosols were pushed away  
 we found that structures form for both bubbles and aerosols in
certain parameter regimes due to the role of the dissipation parameter. 
These regimes can be identified from the bifurcation diagram of the four
dimensional embedding map. 
Both the aerosol and bubble regimes in the phase diagram show regions 
where periodic orbits as well as  chaotic  structures can be seen in the
phase space plots.
In addition to these, fully or partially mixed regimes
can also be seen in the both aerosol and bubble cases.
Thus the dynamic behaviour of the inertial particles can be of  
three major types. 
 
The bifurcation diagrams of the embedding map show further interesting 
structure.
The bifurcation diagrams of the aerosol case show a regime where an
attractor-merging and widening crisis can be seen. 
At the crisis, two period-10 attractors merge and widen simultaneously
into a single chaotic attractor. The parameter values at which the
crisis occurs, are identified using the
Lyapunov exponent and the bifurcation diagram. 
The signature of the attractor
widening crisis can be seen in the intermittency of the time series
of the map variables. The characteristic time before bursts $\tau$ vs
$\gamma_{c} -\gamma$ follows a power-law, $\tau \sim (\gamma_{c}-\gamma)^\beta$
\cite{grebogi87}, where $\gamma_{c}$ is the critical value of the
dissipation parameter $\gamma$ at which the crisis occurs. Thus,
the clustering of advected aerosol particles is dependent
on the dissipation scale.
However, unlike the case of the aerosols, the bubble region did not show
any
of the characteristic signatures of crisis, nor was any intermittency
seen in the bubble region.   

Our results can have implications for practical problems such as the dispersion of 
pollutants by atmospheric flows, and catalytic chemical reactions.

\section{The Embedding Map }

The equation of motion describing the dynamics of passive neutrally
 buoyant particles of finite size  in flows is obtained in \cite{cart02}. 
This equation is called the bailout embedding equation, and is given by,

\begin{eqnarray}
\frac{d}{dt}[\dot{\bf x}-{\bf u}({\bf x})] = - (\lambda I + {\bf \nabla u}) \cdot [\dot{\bf x} - {\bf u}({\bf x})].
\label{neutral}
\end{eqnarray}

The velocity of the particle ${\bf \dot{x}}$ and the local fluid velocity
${\bf u}({\bf x})$  match, when the velocity gradient ${\bf \nabla u}$
exceeds the dissipation $\lambda$, yielding a positive 
$(\lambda I + {\bf \nabla u})$. When $ (\lambda I + {\bf \nabla u}) $ 
becomes negative, the particle will get detached from the
local fluid trajectory and would not follow the fluid. Therefore,
the difference between the particle and the sorrounding fluid velocity
can exponentially grow or get damped depending on
the factor $ (\lambda I + {\bf \nabla u})$.

%The difference between the particle and the sorrounding fluid velocity
%can exponentially grow or get damped depending on
%the factor $ (\lambda I + {\bf \nabla u})$. 
%When the velocity gradient ${\bf \nabla u}$ and the dissipation $\lambda$
%balances yeilding positive $ (\lambda I + {\bf \nabla u}) $ then,
% the velocity of the particle ${\bf \dot{x}}$ will match the 
%local fluid velocity ${\bf u}({\bf x})$. Otherwise, the particle will get 
%detached from the local fluid trajectory and would not follow the fluid.
%From the flow bailout Eqn. \ref{neutral}, the bailout embedding map
%for the neutral particle is obtained in Ref.\cite{cart02} as,

The map analog of the flow bailout in Eqn. \ref{neutral} is obtained in 
Ref \cite{cart02} as,

\begin{eqnarray}
x_{n+2} - T(x_{n+1}) = K(x_{n})[x_{n+1}-T(x_{n})],
\end{eqnarray}
where the base map is an area preserving map given by $x_{n+1}=T(x_{n})$. This is
the bailout embedding map. We note that, the  function $(\lambda I + {\bf \nabla u}) $ in the flow takes the form of  $K(x_{n}) = e^{-\lambda}\partial_{x}T$ in the map case.

For the inertial particle case, the density of the particle and the fluid do not
match. The dynamics of inertial particles in a fluid flow is described by the 
equation \cite{mot03},
\begin{eqnarray}
\frac{d{\bf v}}{dt} - \alpha \frac{d{\bf u}}{dt}  & = & - \gamma({\bf v - u}).	
\label{particle}
\end{eqnarray}	

%This equation is obtained from the Maxey-Riley equations for the transport of a
%spherical particle in a fluid flow  under the approximations discussed 
%in the Appendix \ref{appd} . 
Here, the velocity of the fluid parcel is 
${\bf u}(x,y,t)$, and that of the particle advected by the fluid flow 
is ${\bf v} = d{\bf x}/dt$. The quantity $\alpha$ is the mass ratio parameter 
$\alpha = 3\rho_{f}/(\rho_{f}+2\rho_{p})$, where $\rho_{f}$ is the fluid 
density  and $\rho_{p}$ is the particle density. Thus the $\alpha <1 $  
situation corresponds to aerosols and the $\alpha > 1$ situation corresponds 
to bubbles. The dissipation parameter $\gamma$ is related to $\alpha$ by the
relation $\gamma = 2\alpha/3St$, where $St$ is the Stokes number. The 
dissipation parameter $\gamma$ gives a measure of the expansion or 
contraction in the phase space of the particle's dynamics.

%As the particle dynamics in the fluid flow is compressible, it can 
%be represented by a dissipative map. Let the base flow be represented 
%by  an area preserving map $M$ with the map evolution equation 
%${\bf x}_{n+1} =  {\bf M}({\bf x}_{n})$. The particle dynamics is 
%represented by a map \cite{mot03} which models behaviour 
%analogous to the particle dynamics under%
%Eq. \ref{particle},

Map analog\footnote{Other map analogs of this equation are possible. We follow the form given in Ref. \cite{mot03}} of the flow described by Eq. \ref{particle}
 has been choosen in Ref. \cite{mot03} to  be
,
\begin{eqnarray}
{\bf x}_{n+2} - {\bf M}({\bf x}_{n+1}) &  = & e^{-\gamma} (\alpha {\bf x}_{n+1} - {\bf M}({\bf x}_n))
\end{eqnarray}
where the base map is represented by an area preserving map $x_{n+1}=M(x_{n+1})$.
This can be rewritten,
\begin{eqnarray}
{\bf x}_{n+1} & = & {\bf M(x}_n)+{\bf \delta}_n \nonumber \\
{\bf {\delta}}_{n+1} & = & e^{-\gamma}[\alpha {\bf x}_{n+1}-{\bf M(x}_n)].
\end{eqnarray}

The new variable $\delta$ defines the detachment of the particle from the local 
fluid parcel. This is the bailout embedding map. When the detachment measured by
 the $\delta$ is nonzero, the particle is said to have bailed out of the fluid 
trajectory. In the limit $\gamma \to \infty$ and $\alpha = 1$, ${\bf \delta} 
\to 0$ and the fluid dynamics is recovered. So the fluid dynamics is embedded 
in the particle's equation and is recovered under appropriate limit.  This map
is dissipative with the phase space contraction rate to be $e^{-2a}$.

\subsection{The Embedded Standard Map}

\begin{figure}[!t]
	\begin{center}
	\resizebox{65mm}{65mm}{\includegraphics{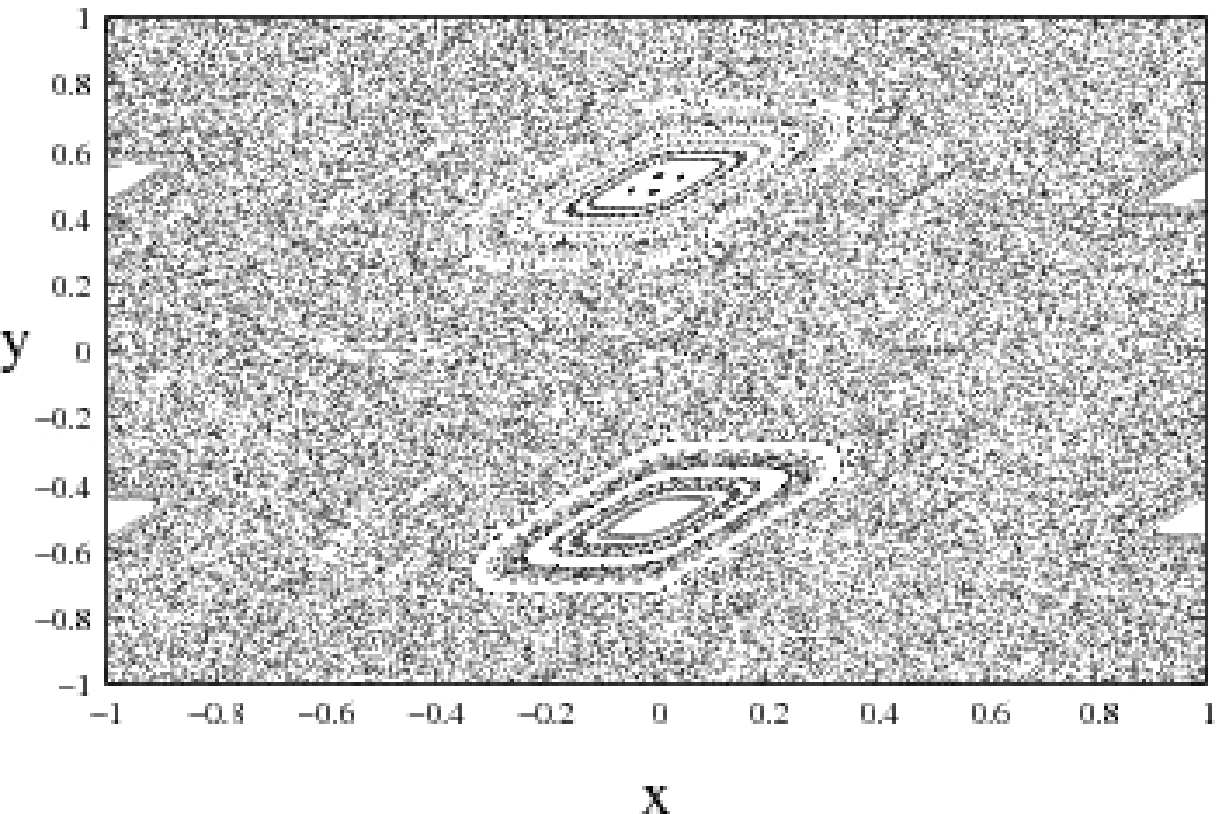}}\\
	\end{center}
   \begin{tabular}{cccc}
(b)&
\hspace{-0.8cm}    \resizebox{65mm}{65mm}{\includegraphics{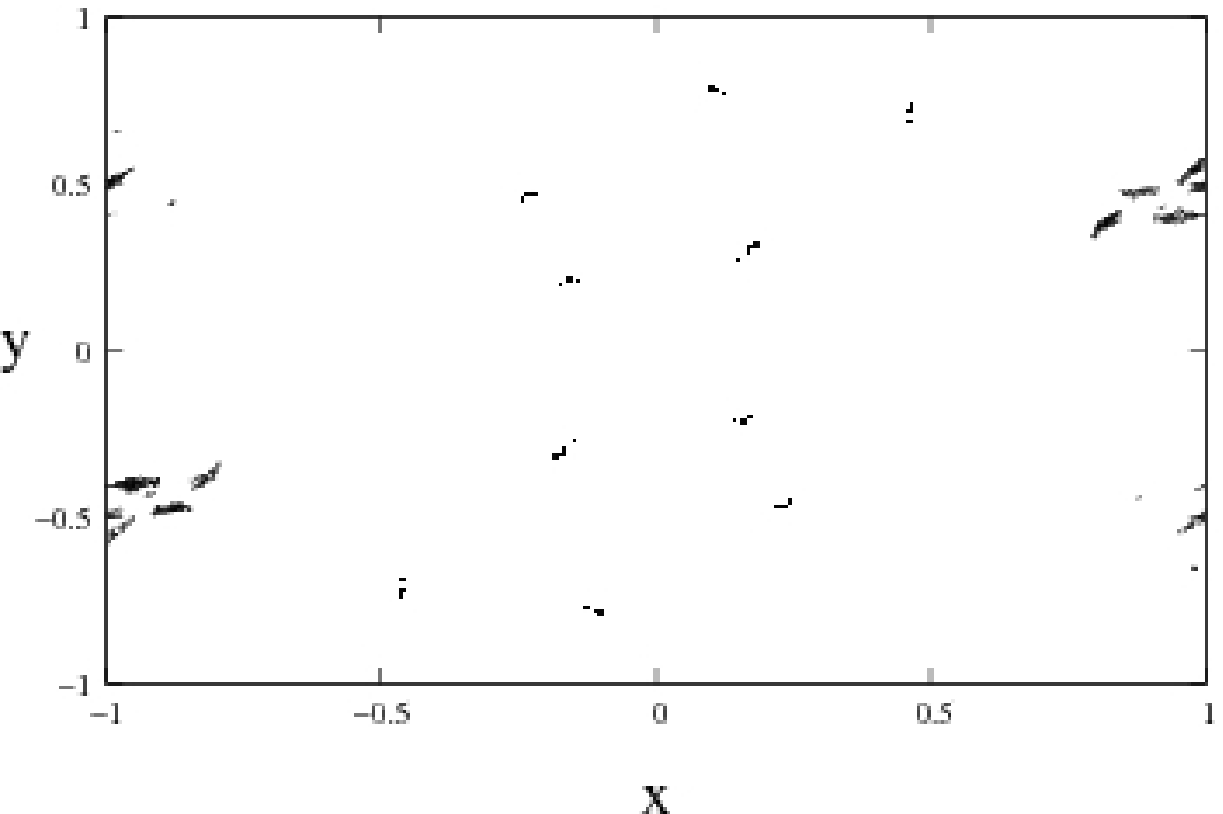}}&
\hspace{0.8cm}(c)&
\resizebox{65mm}{65mm}{\includegraphics{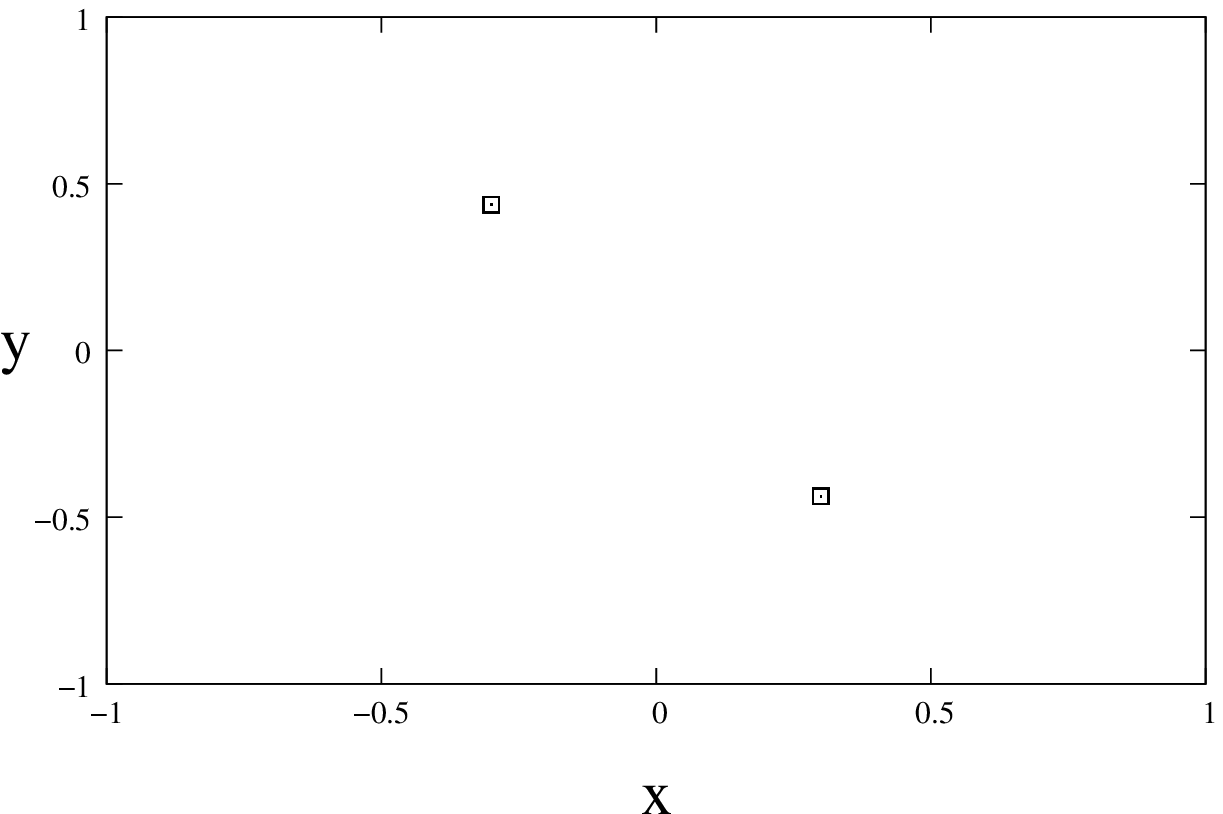}}\\
   \end{tabular}
  \caption{(a) The phase space plot of the standard map for $K=2.0$. (b)
The plot for the corresponding embedding map for aerosols 
($\alpha=0.8$,  $\gamma=0.8$) and (c) bubbles ($\alpha =1.2$,$\gamma = 0.4$).
Random intial conditions distributed uniformly in the phase space were
chosen.} 
\label{phsp-std}
\end{figure}

We choose the standard map \cite{chirikov}, to be our base map as it  
is a prototypical area-preserving system
is widely
studied in a variety of problems of both theoretical and experimental
interest.

The standard map also serves as the usual test bed for the study
of various transport phenomena and their quantifiers\cite{white98}. The
study of impurity dynamics using the standard map as the base flow, can
thus give us a handle on the way in which these phenomena affect the 
impurity dynamics. Earlier studies of impurity dynamics using embeddings
also use the standard map as the base map \cite{cart02}. 

 The parameter $K$ controls the chaoticity of the map. The chaoticity of the
map is dependent on the initial condition and the map nonlinearity  parameter $K$. 
The  map equation is given by,
\begin{eqnarray}
{\bf x}_{n+1} & = & {\bf x}_n + {\bf y}_{n+1} \ ~~~~~~~~~~~ (Mod \  1)\nonumber \\
{\bf y}_{n+1} & = & {\bf y}_n +  \frac{K}{2\pi}\sin( 2\pi{\bf x}_n) \ ~~(Mod \  1)
\label{standard}
\end{eqnarray}

This map is taken to be the base map ${\bf M(x}_n)$ in Eq. (3). The
phase space of this base standard map for $K=2.0$ is plotted in Fig.
\ref{phsp-std} in the region $x \in  [-1,1], y \in   [-1,1]$. 
The particle dynamics is governed by the 4 dimensional map ,
\begin{eqnarray}
x_{n+1} & = & x_{n} + \frac{K}{2\pi}\sin(2\pi y_{n}) + \delta_{n}^{x}  \nonumber\\
y_{n+1} & = & x_{n} + y_{n} + \frac{K}{2\pi}\sin(2\pi y_{n}) + \delta_{n}^{y} \nonumber\\
\delta_{n+1}^{x} & = & e^{-\gamma}[\alpha x_{n+1}-(x_{n+1} - \delta_{n}^{x})] \nonumber\\
\delta_{n+1}^{y} & = & e^{-\gamma}[\alpha y_{n+1}-(y_{n+1} - 	\delta_{n}^{y})]
\label{bailstd}
\end{eqnarray}
It is clear that this  4-dimensional map is  invertible and dissipative.
The  embedding map has 3 parameters $K,\alpha, \gamma$.   
We plot the phase space portrait of the system evolved with random
initial conditions at $K=2.0$, $\gamma=0.5$,
and two values of $\alpha$,  one in the aerosol regime  $0<
\alpha < 1$ (Fig. \ref{phsp-std}(b)), and one in the bubble regime
$1 < \alpha < 3$ (Fig. \ref{phsp-std}(c)). The phase space plot 
of the standard map at the same value of $K$ in Fig. \ref{phsp-std}(a) shows
islands and  chaotic regions separated by invariant tori which
constitute barriers to transport.
We observe from Figs. \ref{phsp-std}(b), and \ref{phsp-std}(c)
that both the aerosols and bubbles have broken the barrier posed 
by the invariant curve and have targetted the  islands forming
structures. Thus, it is clear that 
the dissipation parameter $\gamma$ plays a vital role in the formation
of structures. Therefore in order to identify the regimes in which clustering
or mixing can take place, it is necessary to study the full phase
diagram. The bifurcation diagram of the embedding map can also 
provide insights into the regimes that can be expected in the phase
diagram.  

\begin{figure}[! t]
  \begin{tabular}{cccc}
(a)&
\hspace{-0.8cm}  \resizebox{70mm}{70mm}{\includegraphics{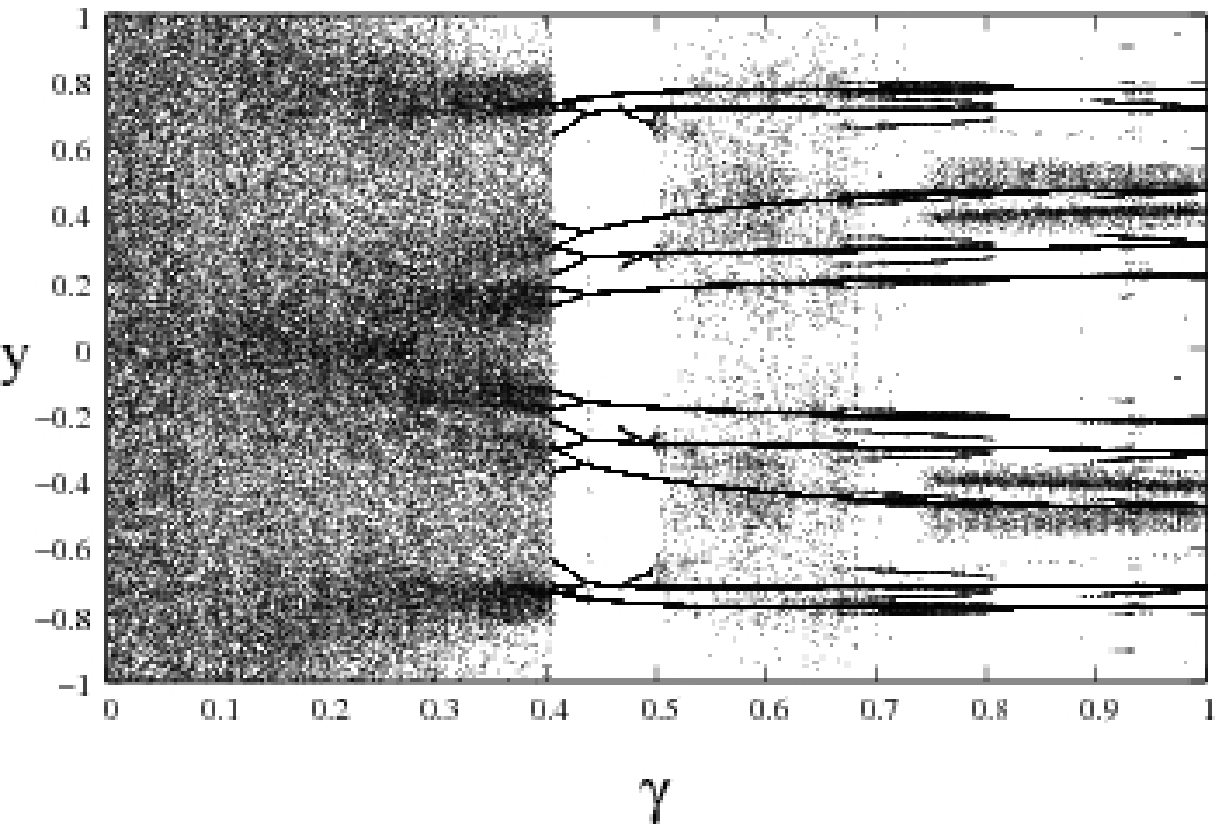}}&
\hspace{0.8cm}(b)&
\resizebox{70mm}{70mm}{\includegraphics{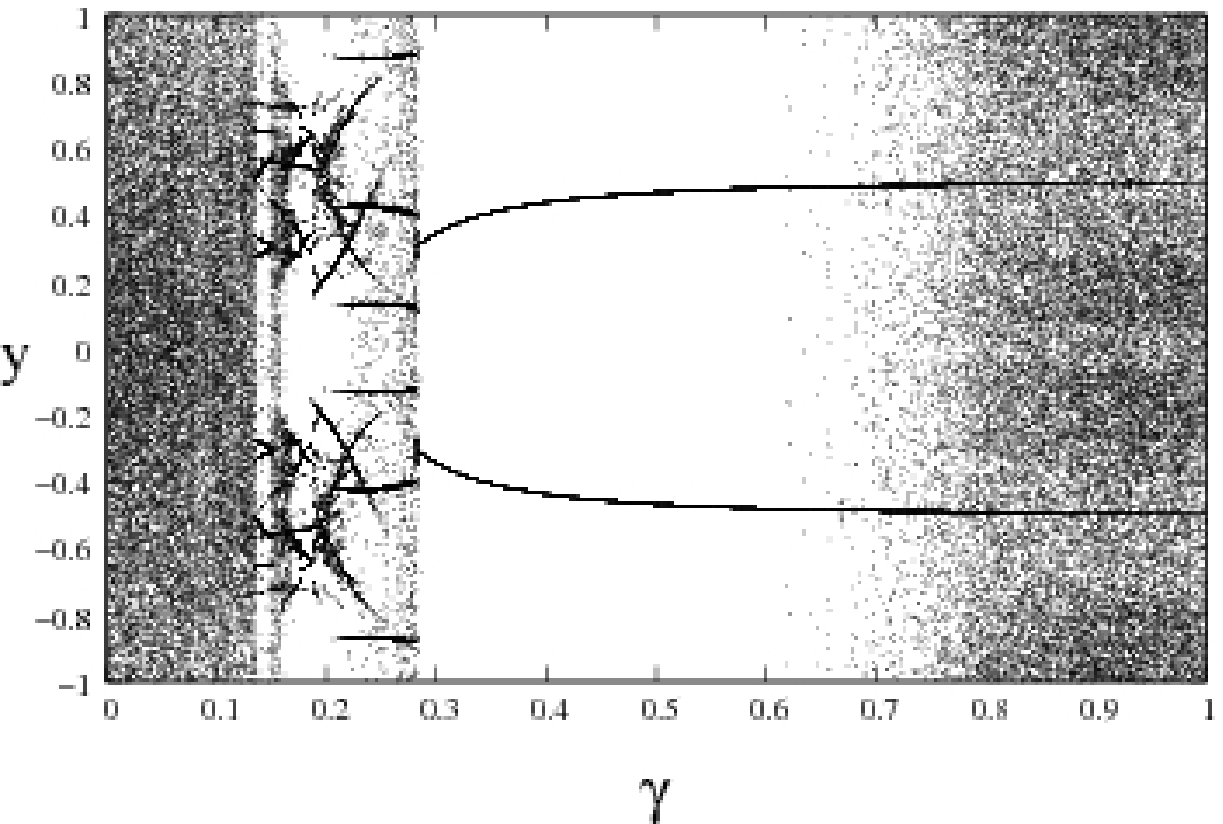}}\\
  \end{tabular}
  \caption{Typical bifurcation diagrams for the embedded standard map for $K =2.0$. 
a) aerosols ($\alpha=0.8$) b) bubbles ($\alpha = 1.2$)  for dissipation parameter
 ranging from $\gamma=0.0$ to $1.0$. \label{bif}}
\end{figure}

%\begin{figure}[! t]
%    \resizebox{90mm}{80mm}{\includegraphics{lp-1p4-dense.eps}}\\
%\caption{The Maximum Lyapunov Exponent plotted versus $\gamma$ for $K=2.0$ and 
%$\alpha =1.4$.\label{lyap-sing}}
%\end{figure}

\section{The Bifurcation Diagram}

\begin{figure}[! t]
  \begin{center}
\resizebox{65mm}{65mm}{\includegraphics{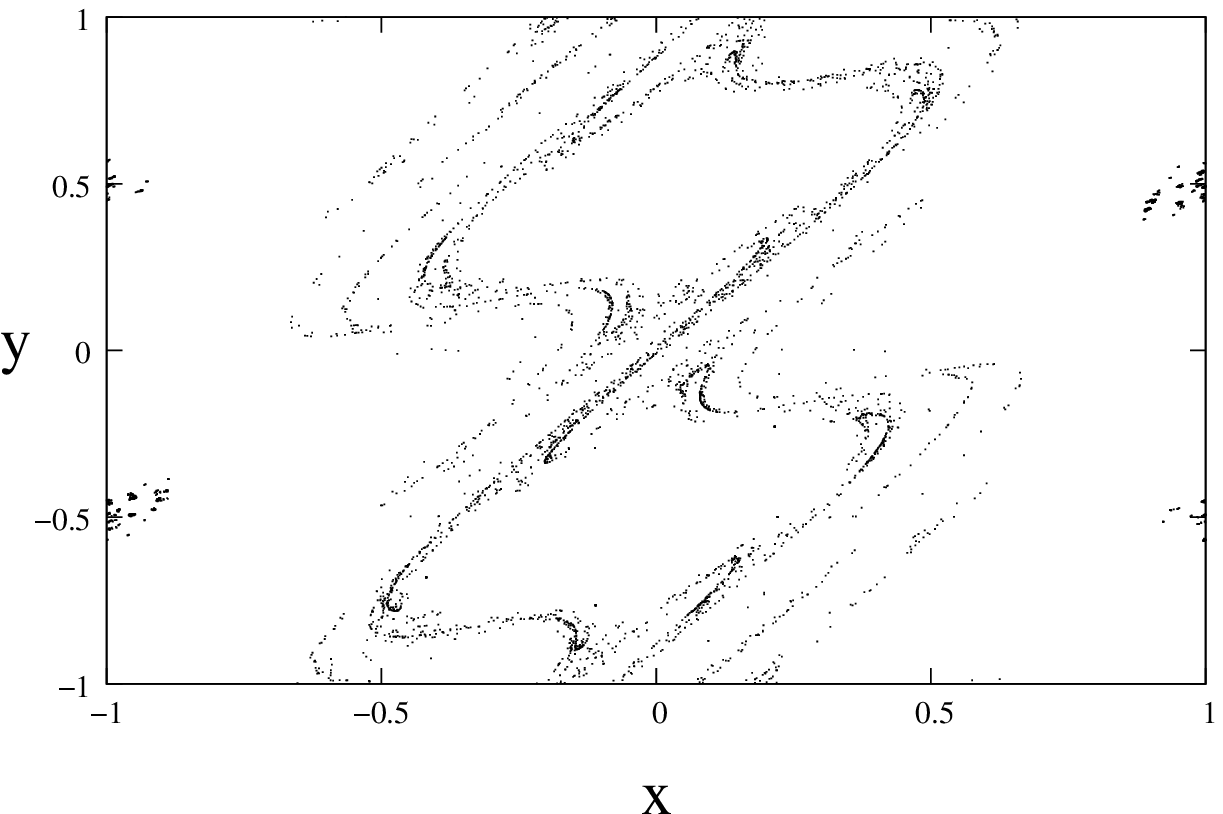}}\\
  \end{center}
\begin{tabular}{ccc}
\hspace{-0.8cm}
 \resizebox{65mm}{65mm}{\includegraphics{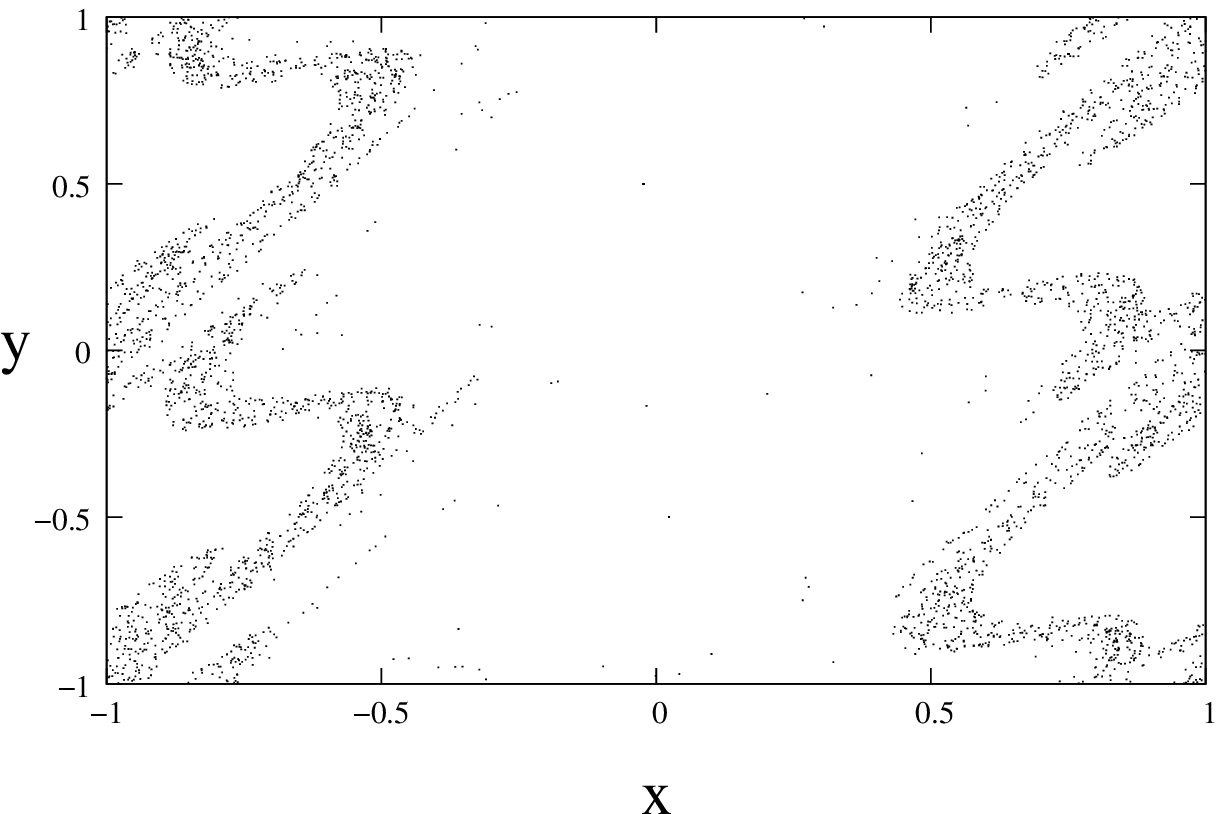}}&
\hspace{0.8cm}
(c)&
\resizebox{65mm}{65mm}{\includegraphics{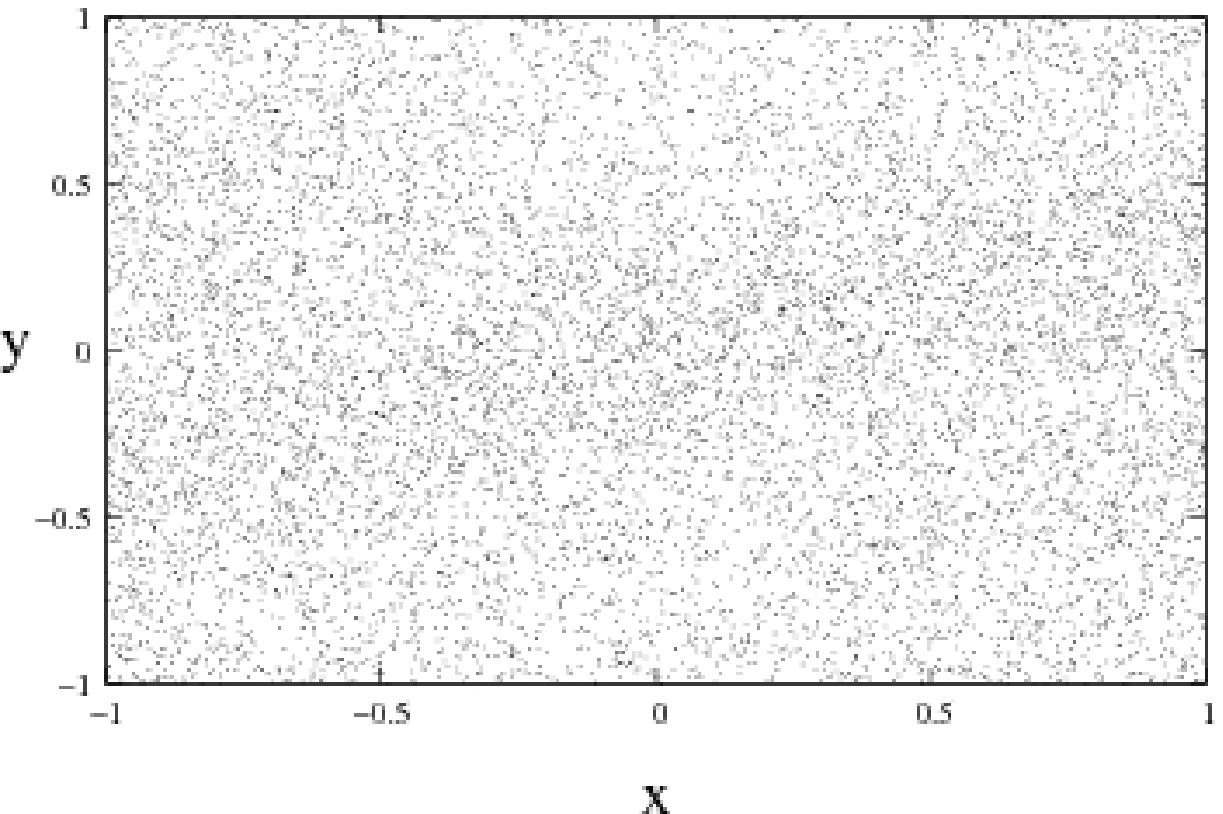}}\\
\end{tabular}                                                 
\caption{Phase space plots for fixed $K=2.0$ showing, (a) a chaotic         
structure in the phase space for the  aerosol case  $\alpha = 0.9$,         
$\gamma = 0.75$; (b) a chaotic structure in phase space for the bubble       
case  $\alpha = 1.05$, $\gamma=0.75$; (c) the mixing region in the
bubble case $\alpha=1.7$, $\gamma=0.3$. \label{pic-psly}}     
\end{figure}

We plot the bifurcation diagram 
of the embedded standard map (Fig. \ref{bif}) for the  parameter values
$K=2.0$ and $\gamma$ values ranging from $0$ to $1$, and for
two $\alpha$ values, $\alpha=0.8$ and $\alpha=1.2$ which 
correspond to the aerosol and bubble cases respectively.
The bifurcation diagrams show many regions of
periodic and chaotic behaviour for both the bubble and aerosol cases. 
This behaviour is reflected in the phase space plots of the map. 
The phase space plots for the aerosol case with
$\gamma=0.8 $, $\gamma=0.3$, and the bubble case with $\gamma= 0.1$ can 
be seen in Fig. \ref{pic-psly}. Regimes with periodic orbits, structures and
mixed regimes can clearly be seen here.

Thus the phase space plots show different types of dynamical behaviour 
depending on the $\gamma$ and $\alpha$ values. The  
phase diagram of the system in $\alpha-\gamma$ space is essential 
for identifying the parameter regimes where different kinds 
of dynamic behaviours can be seen.

\begin{figure}[! t]
\resizebox{100mm}{80mm}{\includegraphics{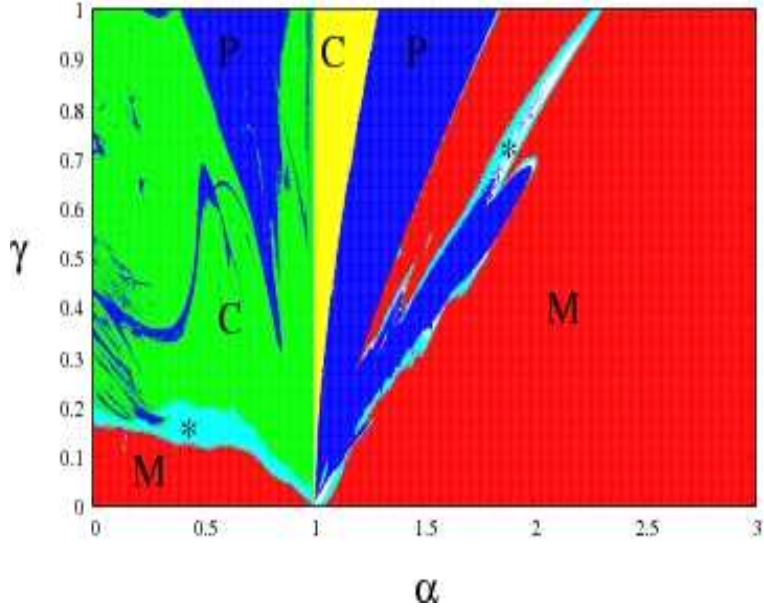}}\\
\caption{(Color Online) The phase diagram.
The regimes with periodic structures are marked
with P (blue). The regimes with chaotic structures are marked with C (green
 for aerosols and yellow for bubbles). The regimes with structures
in a mixing background are marked with a `*'(light blue).
The regimes with full mixing are marked with M (red). \label{lyap-pd}}
\end{figure}

\section{The Phase Diagram}

The  phase diagram of the system can be constructed using the largest
Lyapunov exponent as the characteriser
and is shown in Fig. \ref{lyap-pd}.  For the calculation
of the Lyapunov exponent, the first 5000 iterates are taken as transients 
and the next 1000 iterates are stored, for 100 random initial 
conditions, uniformly spread in the phase space.

The $\alpha$ values range from the aerosol regime $0 \le \alpha < 1$, to the
bubble regime $1<\alpha \le 3$. Tongue-like structures are seen in the
parameter space. Three types of regimes  are seen in the phase diagram 
(i) regimes with periodic structures (P), where the 
largest Lyapunov exponent is  negative, (ii) regimes with
chaotic structures (C), (iii)  mixing regimes
(M), and (iv) chaotic structures in a mixing background.
The Lyapunov exponent takes positive values in regimes (ii),
(iii) and (iv).
In the aerosol regime and the bubble regime, periodic orbits 
like those of Fig. \ref{phsp-std}(b) and (c)
are seen inside the tongues marked by P (blue - (online)),
and chaotic structures like those in Fig. \ref{pic-psly}(b) 
are seen in the regions marked by C (green -(online), in the aerosol
regime and yellow - (online), in the bubble regime) of the phase diagram.
The fully mixing regimes marked with M (red -(online), which exist in both
aerosol and bubble regimes, are identified by the following
procedure. At a given parameter value, the phase space is covered by a 
$100 \times 100$ mesh, and the number of boxes accessed by the iterates is counted. The initial conditions and the transients are same as those that
were used for calculating the Lyapunov exponent. 
The regions of parameter space where the iterates access more than
$99 \%$ of the phase space grid turn out to be  fully mixing regions
with no remanent of any chaotic structure seen in the phase space plot, and have been marked with an M.

\begin{figure}[! t]
\begin{tabular}{cccc}
\hspace{-0.8cm} (a)&
\resizebox{65mm}{65mm}{\includegraphics{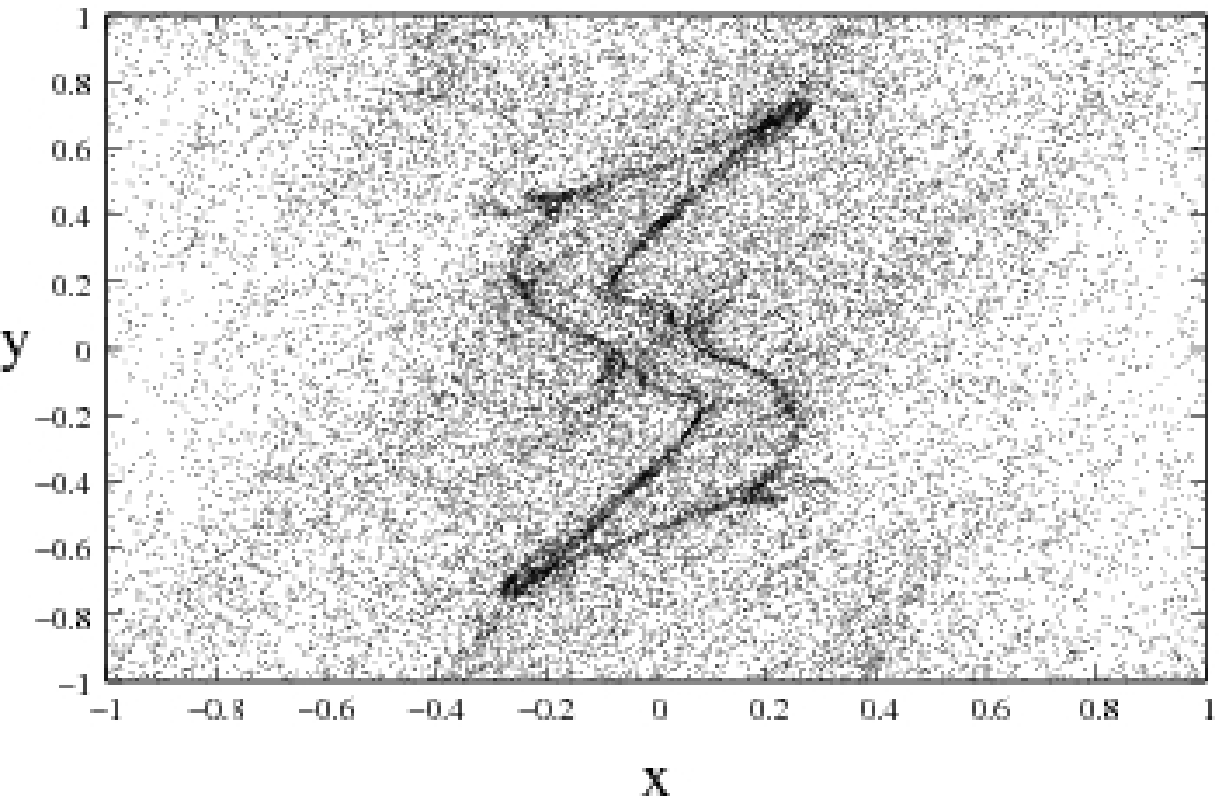}}&
\hspace{0.8cm} (b)&
\resizebox{65mm}{65mm}{\includegraphics{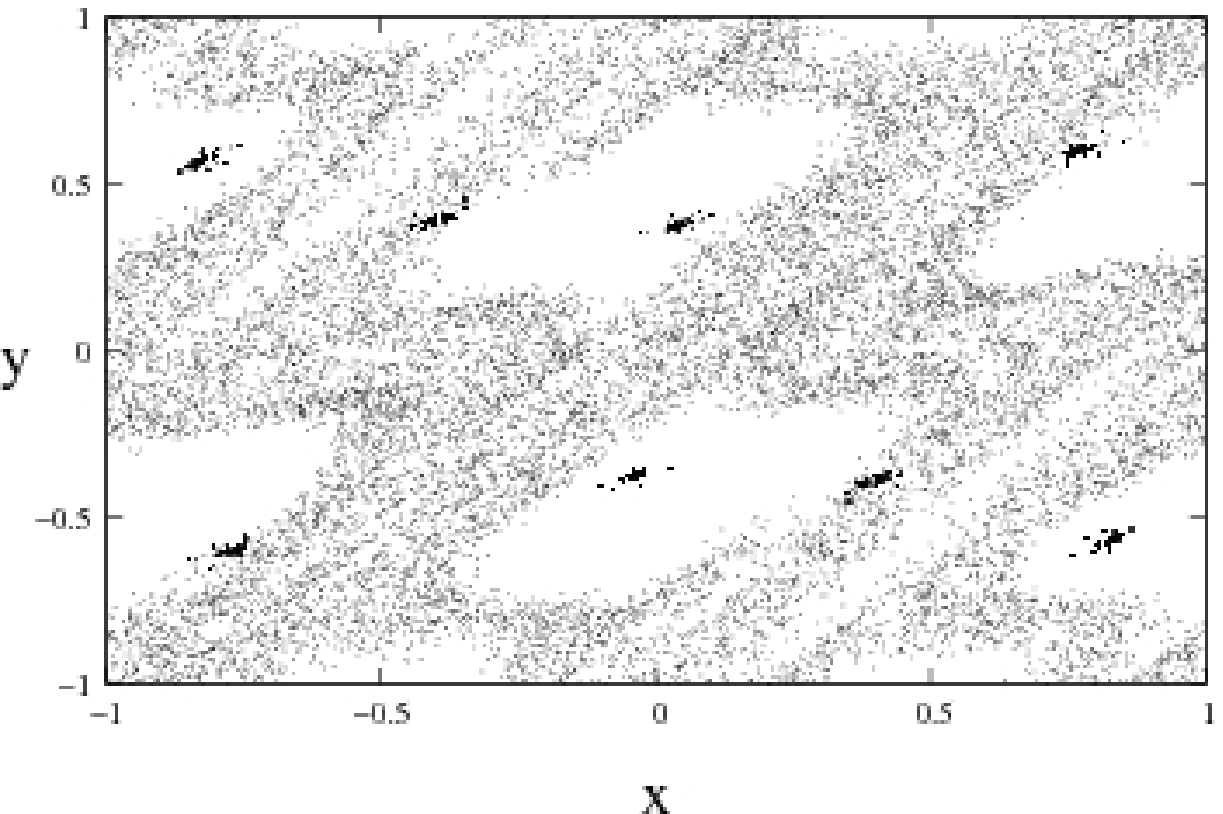}}\\
\end{tabular}
\caption{(a) A typical phase space plot for the regions with occupancy
$<99\%$ and $>90\%$. Chaotic structures are seen in a mixing background.
These regions are identified in the phase diagram - `*' (light blue
-online).(b) A typical phase space plot with occupancy $<90\%$ and $50\%~60\%$. They correspond to the unshaded (blank) regions in the phase diagram.}
\label{unshad-phsp}
\end{figure}

The regions where the iterates  access $>90\%$ and $<99\%$ of the boxes in the phase
space have been marked with an asterisk (*) (light blue-(online)).
The phase space plots corresponding to this region  have
a  chaotic structure in a well mixed background 
(Fig. \ref{unshad-phsp}(a)). 
The tiny patches of unshaded (blank) regions in the phase diagram are regions where 
$\lambda_{max}>0$ and the trajectories access less than $90\%$ 
of the grid in the phase space. Typically,
the trajectories cover $50-60\%$ of the grid 
in the phase space in these regions. Fig. \ref{unshad-phsp}(b)
shows a typical phase space plot for this case.
Here, the iterates  wander densely along the separatrices 
in the phase space, but large voids are seen in 
the phase space. However many iterates stick in 
the neighbourhood of the tiny dense structures seen in the middle 
of these voids.
%end -text added

%Fig. \ref{unshad-phsp}(a) shows a typical phase
%space plot in the unshaded tongue at 
%the bottom of the bubble regime. Here,
%the iterates  wander densely along separatrices 
%in the phase space, but large voids are seen in 
%the phase space. However many iterates stick in 
%the neighbourhood of tiny dense structures in the middle 
%of these voids. Fig. \ref{unshad-phsp}(b) shows a typical plot 
%in the unshaded region in the upper part of the 
%bubble regime. Dense clusters in an otherwise 
%well-mixed regime are seen in Fig. \ref{unshad-phsp}(b). 

\section{The Aerosol case: A crisis in the embedded standard  map}

The bifurcation diagram of the embedded standard map shows regimes where the
attractor undergoes a sudden discontinuous change of type as the system 
parameter is varied.
Such sudden discontinuous changes  signal the onset of a crisis
\cite{grebogi87}. 
The usual crises seen in dissipative dynamics are of the following types.
In the attractor destruction type of crisis, a chaotic attractor is
destroyed as the parameter passes through its critical crisis
value. In the attractor widening or interior crisis, the size of the
chaotic attractor increases suddenly. In the attractor merging crisis,
two or more chaotic attractors merge to form one chaotic attractor. 
It is to be noted that the inverse of the above processes 
(i.e, the sudden creation, shrinking, or splitting of a chaotic attractor) 
occur as the parameter is varied in the other direction. 
In the case of our embedding map, the location and type 
of the crisis can be easily identified from the bifurcation diagram (see
Fig. \ref{bif}(b)).
% A magnified picture of the crisis region can be seen in
%Figure. \ref{pic-bifcri}. It is clear that a widening of the attractor is seen 
%in the system.

%\begin{figure}[!b]
%  \begin{tabular}{c}
%	\resizebox{90mm}{80mm}{\includegraphics{bif-criclose-alp0p8-50p-5t.eps}}\\
%\end{tabular}
%\caption{The bifurcation diagram for $K=2.0$, $\alpha = 0.8$, a close up near $\gamma=0.40$  \label{pic-bifcri}}
%\end{figure}

\begin{figure}[!b]
\begin{tabular}{cccc}
\hspace {-0.8cm} (a) &
\resizebox{70mm}{70mm}{\includegraphics{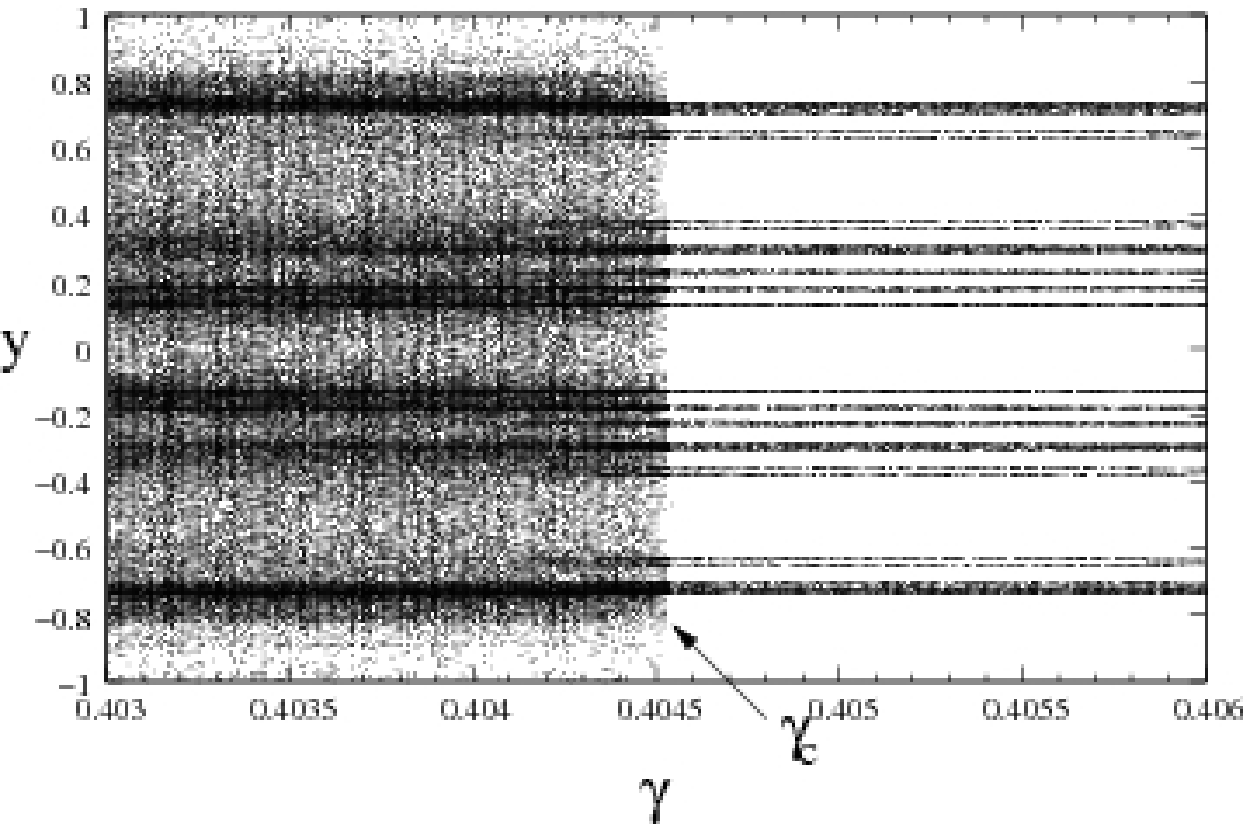}}&
\hspace{0.8cm} (b) &
\resizebox{70mm}{70mm}{\includegraphics{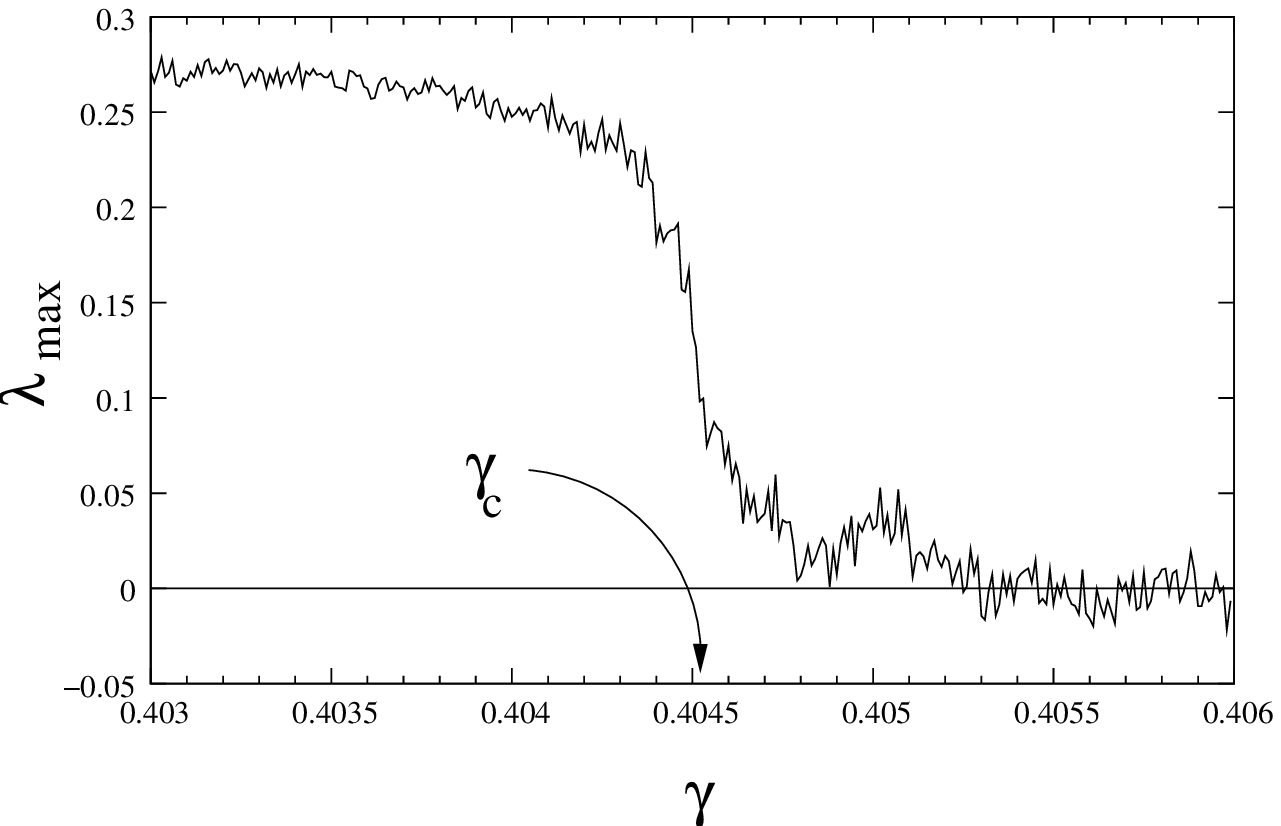}}\\
\end{tabular}
\caption{The identifiers of crisis: (a) the bifurcation diagram (b) the largest Lyapunov exponent plot, both obtained in the vicinity of the crisis, plotted as 
a function of the dissipation parameter $\gamma$ show a sudden change at the 
crisis where the dissipation parameter value $\gamma_{c} = 0.40452$ .  \label{id-cri}}
\end{figure}

       The most prominent change in the size of the attractor is seen
near $\gamma = 0.40$. Here, as the parameter $\gamma$ decreases from 0.41 to 0.40
the attractor widens suddenly. 
The pre-crisis attractor has orbits of finite period, whereas after the
crisis, the trajectories 
access the full range in $y$. The phase space plots at $\gamma
=0.41$ and $\gamma=0.40$, corresponding to the pre-crisis and
post-crisis situation respectively, are seen in Fig. \ref{pre-post}. In
the pre-crisis situation there are two period-10 attractors, 
and trajectories are confined to one or the other attractor, depending
on the initial condition of the trajectory. After the crisis, the two
attractors merge and widen into one.  
Thus we see the  occurence of an attractor merging and widening
crisis. 

The exact parameter value at which the crisis occurs can be identified
using standard quantifiers like the Lyapunov exponents, and 
also the bifurcation diagram.

%\subsection{The identifiers of the crisis}

\begin{figure}[!t]
   \begin{tabular}{cccc}
(a)&
\hspace{-0.80cm}	\resizebox{70mm}{70mm}{\includegraphics{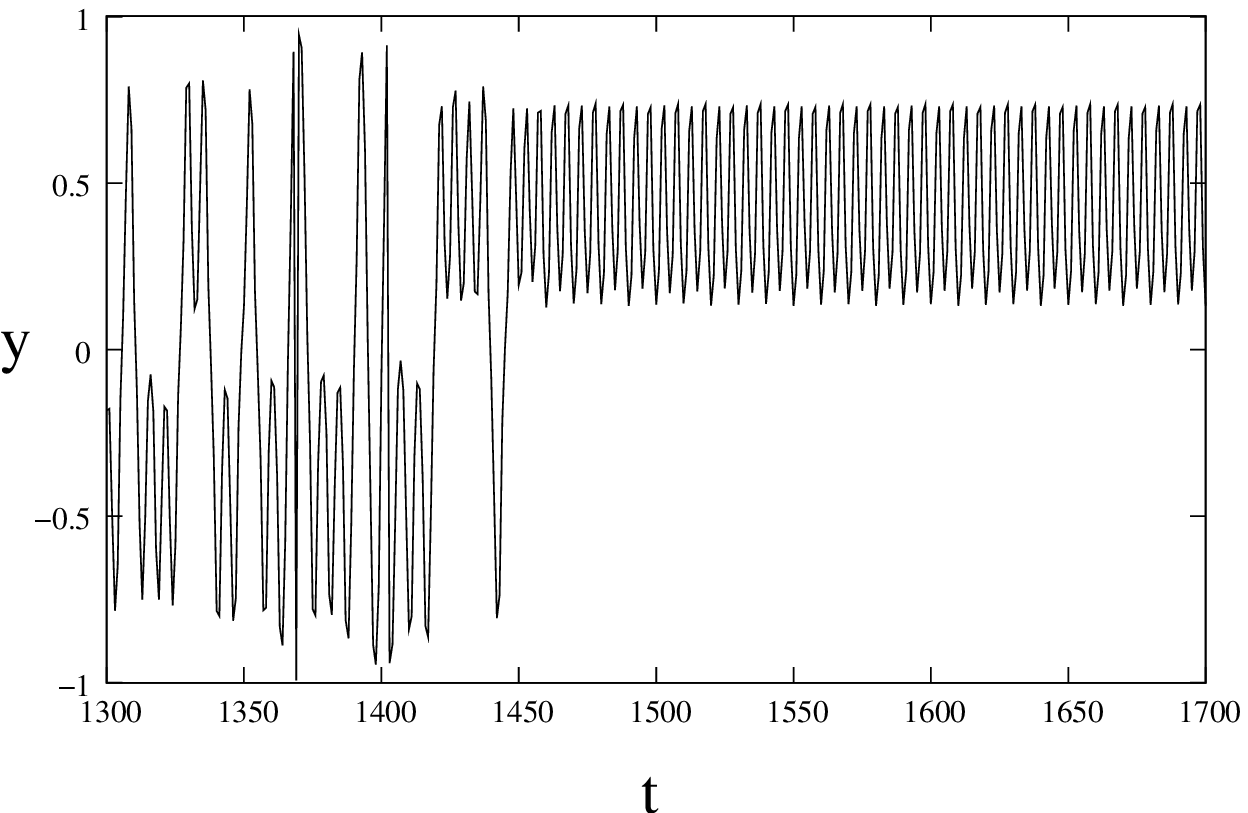}}&
\hspace{0.8cm}(b)&
\resizebox{70mm}{70mm}{\includegraphics{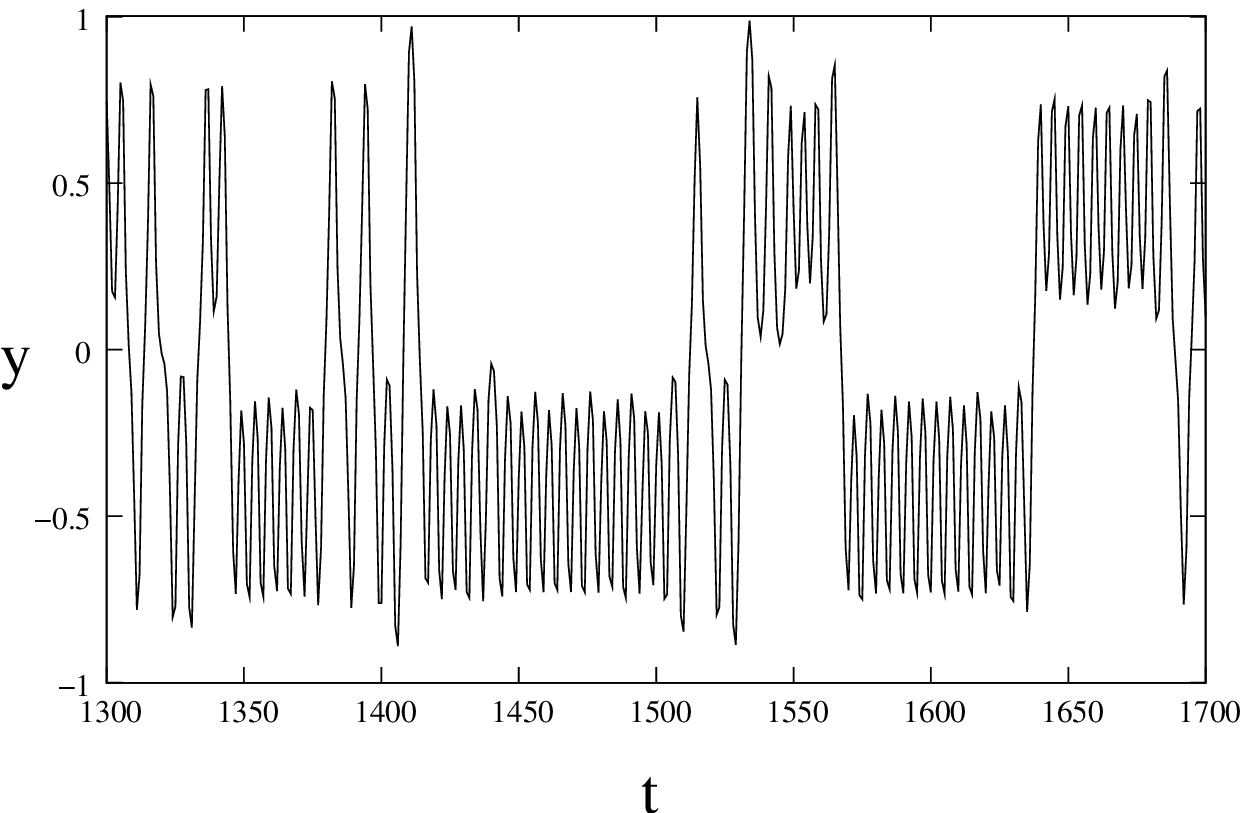}}\\
   \end{tabular}
\caption{The time series plots for a single trajectory  for (a)the
pre-crisis attractor,
 $\gamma = 0.41 $; (b) the post-crisis attractor, $\gamma =0.40$. \label{ts}}
\end{figure}

%The transient time considered is the time taken for an arbitrarily chosen initial
%condition  to settle in either of the attractors. For an
%average over many initial conditions, it is observed that the 
%pre-crisis ($\gamma>0.40452$) transient time is very short and 
%the post-crisis ($\gamma <0.40452$) transient time
%is  arbitrarily long as the system hops between the attractors. The
%transient time is plotted in Figure. \ref{id-cri}(a) for a run of $50,000$ iterates.
The close-up of the bifurcation diagram in Fig. \ref{bif}(b) near $\gamma=0.40$ 
is shown in Fig. \ref{id-cri}(a). As the dissipation parameter $\gamma$
is reduced from $\gamma=0.41$ through $\gamma=0.40$,  a sudden 
widening is seen. The largest Lyapunov exponent, 
can also be used to identify the parameter value at crisis.  
Fig. \ref{id-cri}(b) shows the Lyapunov exponent  plotted in the vicinity 
of the crisis.  The exponent  clearly shows a sudden jump  at $\gamma=0.40452$,
which signifies the occurence of the crisis \cite{vishal}.
From the above identifiers, it is  seen that, the critical
parameter value at crisis is $\gamma_{c}= 0.40452$ 
(to our numerical accuracy).

\subsection{Crisis induced intermittency}

We see the phenomenon of intermittency in the time series of the
phase space co-ordinates in the neighbourhood of the attractor 
widening crisis. Let $\gamma_{c}$ be the parameter value where the attractor
widening crisis occurs. In the pre-crisis region, ($\gamma >\gamma_{c}$),  
trajectories  with arbitrary initial conditions in the
phase space hop between the two attractors for small transient times.
After this brief period, they settle into one of the attractors for
arbitrarily long times (Fig. \ref{ts}(a)).

\begin{figure}[!t]
   \begin{tabular}{cccc}
(a)&
\hspace{-0.80cm} \resizebox{70mm}{70mm}{\includegraphics{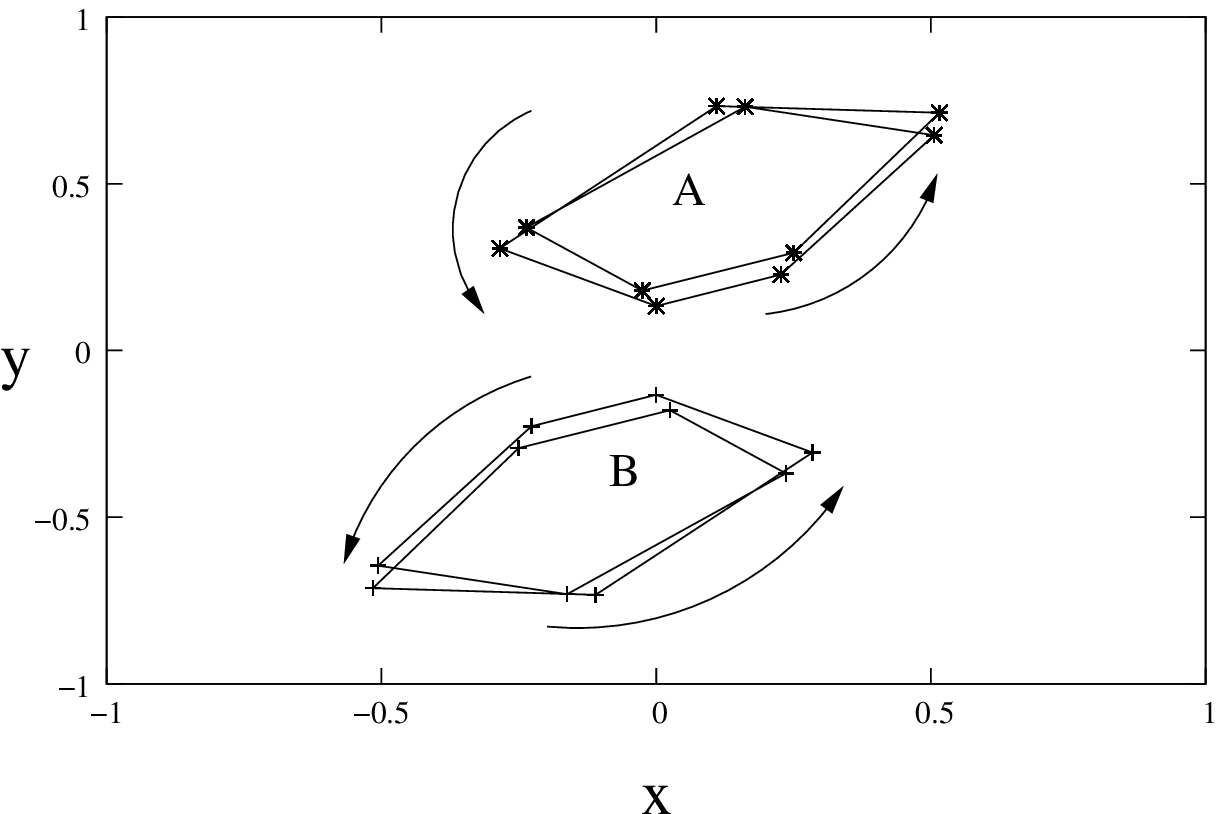}}&
\hspace{0.80cm}
(b)&
\resizebox{70mm}{70mm}{\includegraphics{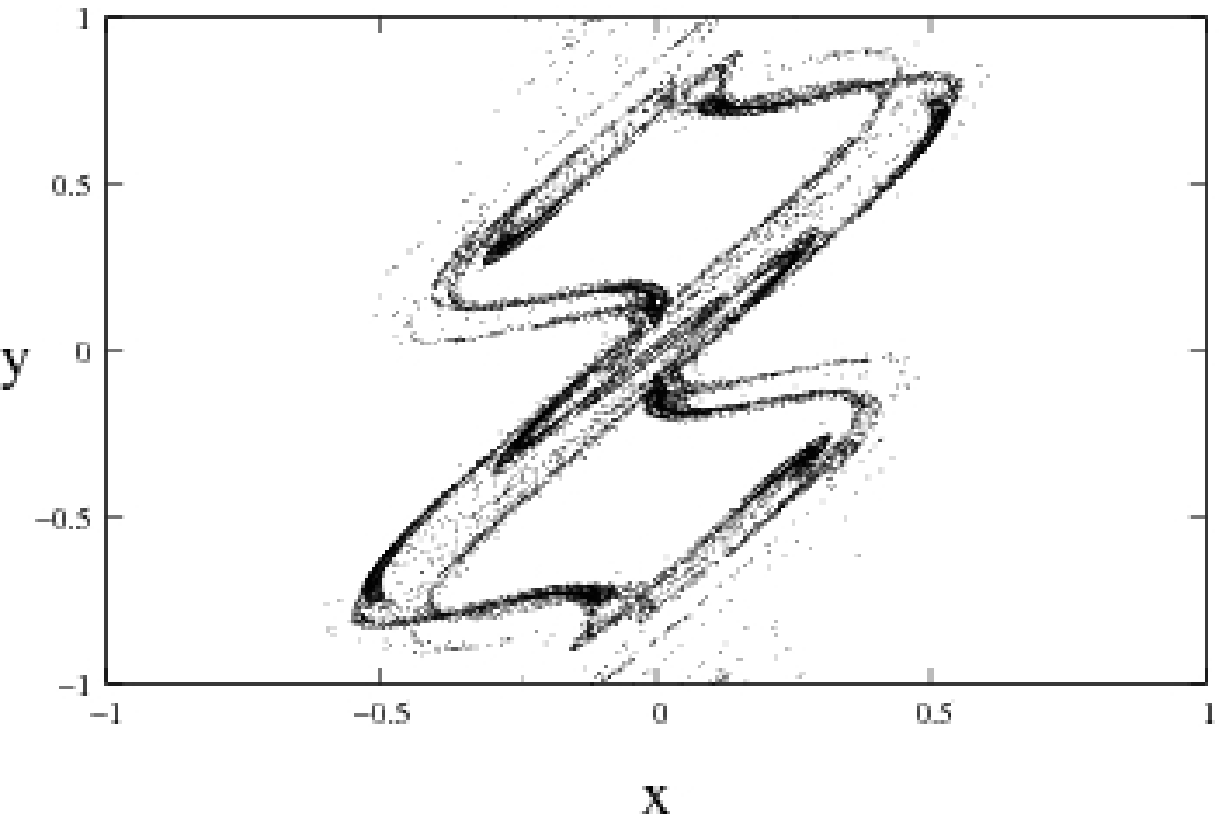}}\\
\end{tabular}
\caption{The phase space plots   (a) the pre-crisis attractor with two period-10 orbits;
 $\gamma = 0.41 $ (b)the post-crisis attractor; $\gamma =0.40$. \label{pre-post}}
\end{figure}

In the post-crisis region, ($\gamma < \gamma_{c}$), trajectories
starting from arbitrary initial conditions hop intermittently between
the regions corresponding to the two pre-crisis
attractors for arbitrarily long times and never settle into either of
the attractors. The time interval between the hopping is random as can
be seen from the Figure. \ref{ts}(b). In the attractor merging
and widening crisis, an orbit initially confined to only one of
the pre-crisis attractors (Figure. \ref{pre-post}(a)), say A or B,
is able to access the widened single  attractor composed of the 
two pre-crisis attractors A and B (Figure. \ref{pre-post}(b)) after
$\gamma$ is reduced to $\gamma_{c}$.

%\subsection{The scaling law for the crisis}

\begin{figure}[!t] \begin{tabular}{c}
        \resizebox{95mm}{72mm}{\includegraphics{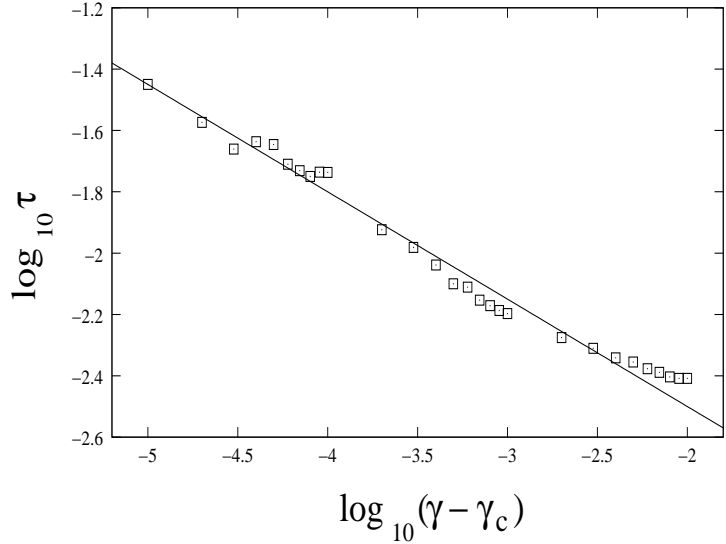}}
\end{tabular}
\caption{The log-log plot of the characteristic time $\tau$  versus $\gamma_{c}-\gamma$ with the slope  $\beta = -0.35 \pm 0.065087$. \label{scale}}
\end{figure}

%The crisis in the system  gives rise to  intermittency. In the post-crisis scenario, 
%the pre-crisis and the larger post-crisis attractors coexist in the phase space and 
%the trajectories hop between the old (pre-crisis) attractor and the new (post-crisis)
%attractor for arbitrarily long times.
When a trajectory leaves the pre-crisis attractor
and accesses the post-crisis attractor, a burst is said to have occurred.
The trajectory settles back into the pre-crisis attractor after a time interval
and a new burst is initiated after some time. The characteristic time
for the orbit to stay in the pre-crisis attractor before a burst occurs
is defined as $\tau$. For each value of $\gamma$ the distribution of the
laminar lengths $\tau$ can thus be obtained. The average characteristic
time $\tau$ in the neighborhood of the $\gamma_{c}$  was seen 
to follow a power law \cite{grebogi87},
\begin{eqnarray}
	\tau \sim (\gamma_{c}-\gamma)^\beta .
\end{eqnarray}

where the exponent was $\beta = - 0.35$. Fig. \ref{scale} 
shows the log-log plot of $\tau$ vs $(\gamma_{c}-\gamma)$ which shows 
this scaling behaviour. 

We note that crisis induced intermittency is seen at several points on the
left edge of the periodic tongue in the aerosol regime of the phase
diagram. Power law scaling behaviour was seen at all these points. 
Unlike the aerosol case, no intermittency or crisis was observed
anywhere in the bubble region. 

 The reason for this difference in behaviour of the aerosols and the
bubbles may lie in the structure of the phase space. The invariant tori
which act as barriers to transport in the neighbourhodd of elliptic
fixed points break up at higher values of $K$ leaving behind leaky
barriers or cantori. In the case of the motion of inertial partcles, 
the elliptic islands and their neighbourhoods act as centrifuges, by
pushing away the heavier aerosols and entrapping the lighter bubbles
\cite{bec07}. At  high values of the dissipation $\gamma$
, the dissipation counteracts the effect of the centrifugal force on the aerosols, so that the they are trapped in the neighbourhood of the elliptic islands, despite
the centrifugal force that pushes them out. However, at lower values of
 dissipation e.g. for our map, at  $\gamma<\gamma_{c}$, where $\gamma_{c}$
 is the criticalparameter at which crisis occurs, the centrifugal force 
pushes out the aerosols through the leaky barries leading to
 attractor widening and intermittency. In the case of the
bubbles, both the dissipation and the centrifugal force work to trap the 
bubbles, and no crisis or intermittency is seen.

As  mentioned above, the standard map which has been studied here is the
prototypical area-preserving map. Hence, we expect that the key result,
viz. that the dissipation of the fluid as well as the buoyancy of the
particles affect the formation of coherent structures and the existence
of mixing regions will apply to generic area preserving system. However, 
the details of the map, such as the existence of Cantori, and the size
of the resonance zones will affect details like the existence of    
phenomena like crisis, crisis induced intermittency etc. 

%However for demonstration we take a quadratic map of the form
%\begin{eqnarray}
%y_{n+1} &=& x_{n}\\
%x_{n+1} &=& 1-y_{n}-a{x^2}_{n}. \nonumber
%\label{quad}
%\end{eqnarray}
%This map has fixed points and chaotic regions sorrounding it. 
%The details of which can be found in \cite{reichl}.
%Here, unlike the standard map, the dissipation could not
% confine the aerosols into any elliptic island. This
%could be attributed due to the
%absence of Cantori in the base map. Though, for
%bubbles the dissipation plays a similar role in trapping them
%for a range of $\gamma$ values and for the rest of $\gamma$
%it is able to access the chaotic by ejecting out of
%the islands. 
 
\section{Conclusion}

To summarize, the present work discusses the advection of passive scalars of finite
size in an incompressible two-dimensional flow for situations where 
the particle density can differ from that of the fluid. 
The motion of the advected particles is represented by an embedding map
with the  area preserving standard map as the base map. 

The new embedded  standard  map is invertible and dissipative and
contains three parameters, the nonlinearity $K$, the dissipation
parameter $\gamma$ and the mass ratio parameter $\alpha$
which measures the extent to
which the density of the particles differs from that of the fluid. 
The phase diagram of the system in the $\alpha-\gamma$ parameter space 
shows very rich structure. Three types of dynamical behaviours are seen 
periodic orbits, chaotic structures, and mixed regimes which can be
partially or fully mixed. 
Periodic structures are seen inside tongues in the
parameter space in the aerosol as well as bubble regimes. 
Thus, the clustering or preferential concentration of inertial 
particles can be seen in both the aerosol and bubble regimes. 
Chaotic structures can be seen outside the tongues 
in the aerosol regime. The bifurcation diagram indicates the 
existence of  crises at parameter values on the edges  of the
tongues in this regime. The crisis is of the attractor merging and
widening type in which two period ten orbits merge and widen into a
chaotic attractor.
Crisis induced intermittency is seen in the aerosol regime with charactersitic 
times $\tau $ which scale as  $(\gamma_{c} -\gamma)^\beta$ where
the exponent $\beta = -0.35$, and $\gamma_c$ is the
critical value at which the crisis occurs. A well mixed regime is also seen for the aerosols.
The bubble regime of the phase diagram shows periodic orbits, chaotic
structures as well as partially and  fully mixed regimes. No crisis or
intermittency is seen here.

The obvious generalisation of our study is to the three dimensional
case volume-preserving case. In this case, due to the presence of
additional degrees of freedom, a much richer phase diagram is expected.
While earlier studies show that regions which target tubular structures
in phase space do exist\cite{cartprl02}, the presence of additional
stretching directions would perhaps indicate much richer structure and 
larger mixing regions in phase space. Detailed studies in this direction
are in progress.

Our results have implications for the preferential concentration 
of inertial particles in flows, as well as for the targetting 
of periodic structures. Earlier results indicated that bubbles could
breach elliptic islands and target structures, whereas aerosols could
not. Our results indicate that the dissipation parameter $\gamma$ plays
as crucial role in this as the density differential parameter $\alpha$,
and the examination of the full phase diagram is necessary to draw
conclusions about the parameter regimes where targetting and breaching
can occur. We also see that mixing regimes can also exist, 
where random initial conditions can spread throughout the phase space.
These results can have implications for the dynamics of impurities in
diverse application contexts, e.g. that of the dispersion of pollutants 
in the atmosphere, flows with suspended microstructures, coagulation of material
particles in flows and catalytic chemical reactions. We hope the present
work will prove to be useful in some of these contexts.

\section{acknowledgement}
N.N.T thanks CSIR, India, for financial support. N.G thanks DST, India, for partial financial 
support under the project SP/S2/HEP/10/2003.

\end{document}